\documentclass[a4paper,11pt]{article}
\pdfoutput=1 

\usepackage{jcappub} 
\usepackage[T1]{fontenc} 
\usepackage{comment}
\usepackage{physics}
\usepackage{todonotes}
\usepackage{hyperref}
\usepackage{subcaption}
\usepackage{dsfont}
\usepackage[normalem]{ulem}
\usepackage{amsmath}
\DeclareMathOperator\arctanh{arctanh}

\title{\boldmath Ricci Reheating Reloaded}
\author[a]{Giorgio Laverda,}
\author[b]{Javier Rubio}

\affiliation[a]{Centro de Astrofísica e Gravitação  - CENTRA,
Departamento de Física, Instituto Superior Técnico - IST,
Universidade de Lisboa - UL, Av. Rovisco Pais 1, 1049-001 Lisboa, Portugal.}
\affiliation[b]{Departamento de Física Teórica and Instituto de Física de Partículas y del Cosmos (IPARCOS-UCM), Universidad Complutense de Madrid, 28040 
Madrid, Spain}

\emailAdd{giorgio.laverda@tecnico.ulisboa.pt}
\emailAdd{javier.rubio@ucm.es}

\abstract{A Hubble-induced phase transition is a natural spontaneous symmetry breaking mechanism allowing for explosive particle production in non-oscillatory models of inflation involving non-minimally coupled spectator fields. In this work, we perform a comprehensive characterisation of this type of transitions as a tachyonic Ricci-heating mechanism, significantly extending previous results in the literature. By performing ${\cal O} (100)$ 3+1-dimensional classical lattice simulations, we explore the parameter space of two exemplary scenarios, numerically determining the main timescales in the process. Based on these results, we formulate a set of parametric equations that offer a practical approach for determining the efficiency of the heating process, the temperature at the onset of radiation domination, and the minimum number of e-folds of inflation needed to resolve the flatness and horizon problems in specific quintessential inflation scenarios. These parametric equations eliminate the need for additional lattice simulations, providing a convenient and efficient method for evaluating these key quantities. } 
\begin{document}
\maketitle
\flushbottom

\section{Introduction}\label{sec:introduction}

The success of inflation is based on its ability to solve the main shortcomings of the hot Big Bang model while providing a mechanism able to generate the primordial density perturbations seeding structure formation. Although numerous theoretical models of inflation have been already excluded by observations, current data sets are still insufficient to identify a particular scenario within the whole inflationary paradigm. Furthermore, there is still a rather limited understanding of how a specific inflationary model should be embedded into a more fundamental particle physics theory. The situation is expected to improve, however, in the next decade, when missions such as the upgraded Simons Observatory \cite{SimonsObservatory:2018koc}, the LiteBird satellite \cite{Hazumi:2019lys, Sugai:2020pjw} or the ground-based CMB Stage 4 program \cite{CMB-S4:2020lpa} will significantly reduce the uncertainties in the tensor-to-scalar ratio \cite{Martin:2014rqa}. In this context, understanding the details of the heating stage following the end of inflation \cite{Bassett:2005xm,Allahverdi:2010xz,Amin:2014eta} becomes of the uttermost importance for providing accurate theoretical predictions to be compared with observations. 

A minimal Hubble-induced heating mechanism based on the time-dependence of the Ricci scalar was recently put forward in the context on non-oscillatory quintessential inflation scenarios, where the absence of a potential minimum gives rise, rather generically, to a kinetic-dominated post-inflationary era \cite{Opferkuch:2019zbd,Bettoni:2021zhq} (cf.~also Refs.~\cite{Figueroa:2016dsc,Nakama:2018gll,Dimopoulos:2018wfg} and Ref.~\cite{Bettoni:2021qfs} for a comprehensive review of quintessential inflation). In this setting, any self-interacting spectator field non-minimally coupled to gravity may potentially undergo a second-order phase transition when the Universe evolves from inflation to kination and the Ricci scalar turns negative. Generally speaking, the transition to the new minima of the potential appearing at large field values involves a tachyonic instability that comes to an end only when the effective mass of the spectator field becomes again globally positive. This symmetry restoration can either take place at the eventual onset of radiation domination or be  effectively induced by the non-linear dynamics following particle production. In particular, if the energy lost by the spectator in each oscillation is smaller than the one associated to the motion of the new Hubble-induced minima as the Universe expands, the field distribution will be able to cross back the origin of the potential \cite{Opferkuch:2019zbd}, giving rise to a bimodal distribution that disappears after a few oscillations \cite{Bettoni:2021zhq}.

The main purpose of this work is to perform a full characterisation of Hubble-induced phase transitions as a tachyonic-heating mechanism, significantly expanding the results in Refs.~\cite{Bettoni:2019dcw, Bettoni:2021zhq}. By performing a large number $\simeq \mathcal{O}(100)$ of 3+1-dimensional classical lattice simulations we will explore a wide portion of the parameter space of the theory. In particular, the outputs of our simulations will allow us to  numerically characterise the main timescales in the process, namely the initial tachyonic instability, the onset of backreaction effects and the virialisation phase leading to a radiation-like equation-of-state for the spectator field. These numerical results constitute the starting point for obtaining a set of ready-to-use parametric formulas characterising the total energy-density at each timescale, the efficiency of the heating process and the radiation temperature at the onset of radiation domination for all the model parameters in the scan, in the spirit of Ref.~\cite{Figueroa:2016wxr}. This line of work comes with several benefits. First, the non-linear phenomenology of Hubble-induced phase transitions is investigated comprehensively, without resorting on \textit{ad hoc} homogeneity assumptions and with a strong emphasis on the interplay between the different model parameters. Second, the simple relations given by the fitting formulas allow us to accurately determine the minimal number of $e$-folds of inflation needed to solve the flatness and horizon problems in specific quintessential inflation scenarios, without resorting to additional time-consuming lattice simulations. Third, the high efficiency of the tachyonic process enable us to directly apply our results to a plethora of variations of the model including, for instance, quartic couplings among species leading potentially to energy transfer effects.

This work is organised as follows. In Section \ref{sec:scalarfield}, we review the basic working assumptions of Hubble-induced phase transitions, focusing for concreteness on a $\mathds{Z}_2$-symmetric model. After deriving analytical solutions for the tachyonic amplification of fluctuations in the linear regime, we proceed to study the non-linear dynamics of the system using
$3+1$ classical lattice simulations. The parametric formulas for the heating efficiency  and the radiation temperature following from this procedure are presented in Section \ref{sec:radtemperature}, where we describe also their impact on the main inflationary observables and the potential effects of non-gravitational interactions with other beyond the Standard Model sectors. Finally, our conclusions are presented in Section \ref{sec:conclusion}.

\section{Spectator field dynamics}\label{sec:scalarfield}
The simplest realisation of a Hubble-induced phase transition is described by an action containing an energetically-dominant quintessential scalar field $\phi$ \cite{Peebles:1998qn, Bettoni:2021qfs, deHaro:2021swo} and a subdominant $\mathds{Z}_2$-symmetric spectator field $\chi$ non-minimally coupled to gravity,
\begin{equation}
    S=\int d^4x \sqrt{-g} \left[ \frac{M^2_P}{2}R + \mathcal{L}_{\phi}  -\frac12{\partial_{\mu}}{\chi} \partial^{\mu} \chi - \frac12 \xi R \chi^2 - \frac{\lambda}{4}\chi^4 \right] \,,   \label{eq.lagrangian_chi_scalar}
\end{equation}
with $R$ the Ricci scalar and $M_{P}=(8\pi G)^{1/2}=2.4\times 10^{18}\, {\rm GeV}$ the reduced Planck mass. The precise form of the inflaton Lagrangian density ${\cal L}_\phi$ in this expression will not play a relevant role in the following discussion, being the associated scalar field only responsible for driving the background evolution of the Universe at early times and, most importantly, for inducing a transition from inflation to a kinetic-dominated epoch. Regarding the spectator field counterpart, it includes a dimensionless coupling constant $\xi$, only restricted \textit{a priori} by the self-consistency of the procedure. In particular, the field-dependent contribution to the overall Ricci scalar prefactor is required to be subdominant as compared to the usual Planck mass counterpart, namely $\xi \chi^2\ll M^2_P$. In this limit, the graviton propagator is well-defined for all field values of interest, while coinciding approximately with the graviton propagator in General Relativity. Moreover, in order to avoid the formation of large isocurvature perturbations, the strength of the non-minimal coupling is restricted to exceed $1/12$, guaranteeing the spectator field to be sufficiently massive as not to undergo large displacements during inflation \cite{Bettoni:2021zhq, Bettoni:2018utf, Opferkuch:2019zbd}. As it will become clear in Sections \ref{sec:linear_regime} and \ref{sec:lattice}, a more restrictive lower bound comes from minimizing the effects caused by reentering modes close to the Hubble horizon, which enforces $\xi\gtrsim15$. In this parameter range, inflationary isocurvature perturbations are safely under control.

Finally, note that, although the phenomenology of Hubble-induced phase transitions can be easily extended to arbitrary monomial potentials \cite{Bettoni:2019dcw, Bettoni:2021qfs},  we have intentionally restricted ourselves to a marginal quartic self-interaction with dimensionless coupling constant $\lambda$. This avoids the introduction of additional energy scales beyond the constant Planck mass and the dynamical scalar curvature entering the Einstein-Hilbert term.

In a Friedmann-Lemaître-Robertson-Walker background $g_{\mu \nu}=\text{diag}(-1, \, a^2(t)\delta_{ij})$ with scale factor $a(t)$, the Klein-Gordon evolution equation following from the variation of Eq.~\eqref{eq.lagrangian_chi_scalar} with respect to $\chi$ takes the form 
\begin{equation}
    \ddot{\chi} + 3H\dot{\chi} - \frac{1}{a^2} \nabla^2 \chi + \xi R\chi + \lambda \chi^3 = 0 \, ,
    \label{eq:eomchi_scalar}
\end{equation}
where dots denote derivatives with respect to the coordinate $t$. The energetic subdominance of the spectator field allows us to interpret the Ricci scalar in this expression as an effective time-dependent mass for $\chi$, whose dynamics in a Friedmann-Lemaître-Robertson-Walker metric is completely determined by the evolution of the inflaton field, namely 
\begin{equation}
    R=6\left(\dot{H} + 2H^2 \right) = 3(1-3w_\phi)H^2\,, \label{eq:ricciscalar}
\end{equation}
with $H = \dot{a}/a$ the Hubble rate and $w_\phi$ the effective inflaton equation-of-state parameter. In these expressions, the Ricci scalar constitutes a cosmic timekeeping device, changing sign precisely at the transition between inflation ($w_\phi=-1$) and kination ($w_\phi=1$) and triggering with it the spontaneous breaking of the $\mathds{Z}_2$ symmetry in the $\chi$ sector of the theory. Given the time-dependence of the resulting symmetry-breaking potential, the true vacua appearing at large field values migrate monotonically towards zero as the Hubble function decreases with time. 

In a generic quintessential-inflation scenario, the transition from inflation to kination takes place on a timescale dictated by the detailed shape of the inflationary potential at the end of inflation. Since we are mainly interested in the production of particles upon spontaneous symmetry breaking, we will assume the crossover stage between these two cosmological epochs to be essentially instantaneous, neglecting with it any particle creation effects prior to the onset of kinetic domination. This assumption allows us to parametrise the evolution of the scale factor and the Hubble rate as
\begin{equation}
    a(t)=a_{\rm kin}[1+3H_{\rm kin}(t-t_{\rm kin})]^{1/3} \, ,\hspace{10mm}
    H(t)=\frac{H_{\rm kin}}{1+3H_{\rm kin}(t-t_{\rm kin})} \, ,
    \label{eq:paramexpansion}
\end{equation} 
with  $a_{\rm kin}=a(t_{\rm kin})$ and $H_{\rm kin}=H(t_{\rm kin})$ the associated values at the beginning of kination. In order to remove the Hubble-friction term in Eq.~\eqref{eq:eomchi_scalar}, it is also convenient to transform the field and spacetime coordinates to their conformal version 
\begin{equation}
 Y = \frac{a}{a_{\rm kin}}\frac{\chi}{\chi_{\rm kin}} \,,  \hspace{10mm}
 \mathbf{y} = a_{\rm kin}\chi_{\rm kin}\mathbf{x} \, ,\hspace{10mm} 
    z = a_{\rm kin}\chi_{\rm kin}\tau \,, 
    \label{eq:variablesredef_scalar}
\end{equation}
where $\tau$ is the usual conformal time variable ($d\tau =dt/a$) and $\chi_{\rm kin} = \sqrt{6 \xi}H_{\rm kin}$ is a typical field value associated to the initial curvature of the effective potential around the origin. Taking into account all these changes, the final version of the Klein-Gordon equation for the scalar field $Y$ takes the form
\begin{equation}
    Y'' - \nabla^2 Y - M^2(z)Y + \lambda Y^3 = 0\,,
    \label{eq:Yeom}
\end{equation}
with 
\begin{equation}
    M^2(z) = \frac{\nu^2 - 1/4}{(z+\nu)^2}=(4\nu^2-1){\cal H}^2 
    \label{eq:time_dependent_mass}
\end{equation}
 an effective time-dependent mass term and
\begin{equation} 
\nu = \sqrt{\frac{3\xi}{2}} \,,  \hspace{10mm}  
\mathcal{H}(z)= \frac{a'}{a}=\frac{1}{2(z+\nu)} \,, \hspace{10mm}  a(z)=a_{\rm kin}\sqrt{1+\frac{z}{\nu}}\, .
\end{equation}
From Eq.~\eqref{eq.lagrangian_chi_scalar}, we can additionally compute the spectator energy-momentum tensor and, in particular, its energy density and pressure \cite{Bettoni:2021zhq}
\begin{eqnarray}
    \rho_{\chi}(\tau) &=&  \frac{1}{2} \chi'^2 + \frac{1}{2a} (\nabla\chi)^2 + \frac{\lambda}{4} \chi^4 + 3\xi\mathcal{H}\chi^2 + 6\xi\mathcal{H}\chi\chi' - \frac{\xi}{a^2}\nabla^2\chi^2 \,,      \label{eq:chi_energy_density_pressure_scalar0} \\
    p_{\chi}(\tau) &=& \frac{1}{2}(1-4\xi) \chi'^2 - \frac{1-12\xi}{6a^2} (\nabla\chi)^2 - \frac{\lambda}{4} \chi^4 + 2\xi\mathcal{H} \chi\chi' \nonumber \\ 
    &-& \frac{\xi}{3a^2} \nabla^2\chi^2 + 2\xi \lambda \chi^4 + \xi \left[\mathcal{H}^2 +12\left( \xi -\frac16\right) \left(\frac{a''}{a} + \mathcal{H}^2 \right) \right]\chi^2  \,.
    \label{eq:chi_energy_density_pressure_scalar}
\end{eqnarray}
Note that, due to the non-minimal coupling to gravity, these quantities turn out to depend non-quadratically on the spectator field value and its velocity, being therefore not guaranteed to be positive definite at all times \cite{Ford:1987de,Ford:2000xg,Bekenstein:1975ww,Flanagan:1996gw}. Although this well-known fact leads to the violation of the weak energy condition within the spectator field sector, the total energy density of the Universe, i.e.~that including the dominant inflaton contribution, is still positive definite at all times.

\subsection{Linear regime} \label{sec:linear_regime}

At this stage, one could be tempted to treat the spectator field $\chi$ as an homogeneous component akin to the runaway inflaton field $\phi$, reducing the analysis of Hubble-induced phase transitions to the mere integration of the homogeneous equations of motion, as done for instance in Refs.~\cite{Figueroa:2016dsc,Opferkuch:2019zbd, Dimopoulos:2018wfg}. Despite its simplicity, this naive approach is based on the false premise that the only relevant mode for the spectator field's evolution is the zero mode. On the contrary, as argued in Refs.~\cite{Bettoni:2018utf,Bettoni:2018pbl,Bettoni:2019dcw} and shown explicitly in Ref.~\cite{Bettoni:2021zhq}, the existence of spatial gradients plays a crucial role in the formation of topological defects, in the overall energy budget and in determining the effective equation of state of the spectator field (cf. also Refs.~~\cite{Felder:2001kt, Felder:2000hj} and \cite{Lozanov:2016hid,Lozanov:2017hjm} for previous analyses supporting this view). Moreover, the homogeneous-field approximation does not respect the fundamental $\mathds{Z}_2$ symmetry of the model under consideration, requiring \textit{ad hoc} initial conditions that cannot be sourced by stochastic quantum fluctuations after inflation comes to an end. 

Having in mind the above arguments, let us proceed to the analysis of the spectator field fluctuations. To this end,  we notice that immediately after symmetry breaking the non-linear term in the evolution equation \eqref{eq:Yeom} plays a rather subdominant role, allowing us to approximate this expression by that for a free scalar field with time dependent mass $M(z)$, making therefore the problem exactly solvable. As described in detail in Refs.~\cite{Bettoni:2019dcw, Bettoni:2021qfs, Bettoni:2021zhq}, the associated spectator field quantisation proceeds according to the standard procedure in non-trivial backgrounds \cite{Birrell:1982ix,Mukhanov:2007zz}. In particular, each Fourier mode $\boldsymbol{\kappa} =\mathbf{k}/(a_{\rm kin}\chi_{\rm kin})$ of the scalar field $Y$ is promoted to a quantum operator
\begin{equation}
\hat Y(\boldsymbol{\kappa},z)=f_{\kappa}(z) \hat a(\boldsymbol{\kappa},z_0) + f_{\kappa}(z)^{\ast} \hat a^{\dagger}(-\boldsymbol{\kappa},z_0)
\end{equation}
that satisfies the equal-time commutation relations $[\hat Y_{\boldsymbol{\kappa}}(z),\hat \Pi_{\boldsymbol{\kappa}'}(z)] = i\delta(\boldsymbol{\kappa}+\boldsymbol{\kappa}')$, where ${\hat \Pi_{\boldsymbol{\kappa}} = d\hat Y_{\boldsymbol{\kappa}}/dz}$, $\hat a_{\boldsymbol{\kappa}}$ and $\hat a^\dagger_{\boldsymbol{\kappa}}$ are the standard annihilation and creation operators and $f_\kappa$ is a set of isotropic mode functions fulfilling the Klein-Gordon equation
\begin{equation}
    f_{\kappa}''(z) + (\kappa^2-M^2(z))f_{\kappa}(z)=0\,.
    \label{eq:mode_eq_chi}
\end{equation} 
To solve this initial-value problem, we will assume the $\nu$ parameter within the effective square mass $M^2(z)$ to be large enough as to sufficiently separate the scales corresponding to the Hubble horizon and the typical momenta amplified by the tachyonic instability,
\begin{equation}\label{tachband}
\mathcal{H} \leq \kappa \leq \sqrt{4\nu^2 - 1}\mathcal{H}\,,
\end{equation} 
thus reducing potential horizon-reentry effects \cite{Bettoni:2019dcw}. Given this requirement and assuming vacuum initial conditions $f_{\kappa}(z_0)=1/\sqrt{2\kappa}$ and $f'_{\kappa}(z_0)=-i\sqrt{\kappa/2}$, the Klein-Gordon equation \eqref{eq:mode_eq_chi} admits a solution \cite{Bettoni:2019dcw}
\begin{equation}
    f_{\kappa}(z)=\sqrt{z+\nu}\left[\alpha_{\kappa} \mathcal{J}_{\nu}(\kappa(z+\nu)) - \beta_{\kappa} \mathcal{Y}_{\nu}(\kappa(z+\nu))\right]\,,
     \label{eq:solY}
\end{equation}
with $ \mathcal{J}_{\nu}$ and $ \mathcal{Y}_{\nu}$ the Bessel functions of the first kind and the quantities $\alpha_{\kappa}$, $\beta_{\kappa}$, $\delta$ defined as 
\begin{align}
    \alpha_{\kappa}    &= \mathcal{Y}_{\nu}(\kappa\nu)\delta \, , \hspace{10mm}
    \beta_{\kappa}     = \mathcal{J}_{\nu}(\kappa\nu)\delta - \frac{f(0,\kappa)}{\sqrt{\nu}\mathcal{Y}_{\nu}(\kappa\nu)} \, ,\\
    \delta      &=\frac{\pi f(0,\kappa)}{4\sqrt{\nu}} \left[1 - 2\nu + 2\nu \left(\kappa\frac{\mathcal{Y}_{\nu-1}(\kappa\nu)}{\mathcal{Y}_{\nu}(\kappa\nu)} - \frac{f'(0,\kappa)}{f(0,\kappa)}\right)  \right] \, .
\end{align}
This solution consists of a rapidly growing mode ($\mathcal{J}$) and a decaying mode ($\mathcal{Y}$). By considering only the former, one can easily obtain the following approximated expression for the amplitude of the spectator field mode functions in the large-$\nu$ limit \cite{Bettoni:2019dcw},
\begin{equation}
    |f_{\kappa}(z)|^2 \approx \frac{1}{\kappa} \left(\frac{2 \nu -1}{4 \sqrt{2} \nu }\right)^2 \left(\frac{z}{\nu }+1\right)^{2 \nu +1} \exp \left[-\frac{\kappa^2 (\nu +z)^2}{2 (\nu +1)}\right] \, \,, 
    \label{eq:fk2}
\end{equation}
which, for highly-symmetric spherical configurations, leads to a two-point correlation function 
\begin{equation}\label{eq:size0}
\zeta(r,z) = \int \frac{d^3\kappa }{(2\pi)^3} \, e^{i\boldsymbol{\kappa} \cdot \boldsymbol{y}}\, \vert f_\kappa(z)\vert^2 \simeq \zeta(0,z)  \,G_1(\kappa_* r) \,,
\end{equation}
with root mean-square perturbation (rms)
\begin{equation}\label{eq:y2}
 \zeta(0,z)\equiv Y^2_{\rm rms}(z)= \left(\frac{2 \nu-1}{8\pi \nu}\right)^2    \left(1+\frac{z}{\nu}\right)^{2 \nu +1} \kappa_*^2(z)\,,
\end{equation}
and shape function
\begin{equation}\label{eq:shape}
G_1(\kappa_* r)\equiv \sqrt{\frac{\pi}{2}}\frac{1}{\kappa_* r} \exp\left(-\frac{1}{2}\kappa^2_* r^2\right) \text{erfi}\left(\frac{ \kappa_* r}{\sqrt{2}}\right)\,.
\end{equation}
Here $\text{erfi}$ stands for the imaginary error function and $\kappa_*(z)\equiv 2\sqrt{\nu+1}\, \kappa_{\rm min}(z)$ is a typical momentum scale. 

The analysis of the system in its linear regime gives us the time-evolution of the amplified scalar modes during the initial tachyonic instability. Their exponential growth leads to a rapid enhancement of the mode occupation number and to the classicalisation of the system  \cite{Bettoni:2021zhq, Bettoni:2021qfs, Bettoni:2019dcw}. We will use this result for two different purposes. Firstly, it allows us to estimate the typical momenta at play, namely the lowest and highest momenta in the tachyonic band (see Eq.~\eqref{tachband}) and the typical momentum $\kappa_*$. This information is crucial when we perform numerical lattice simulations of the system in its full non-linear glory. Secondly, we can take advantage of the solution in Eq.~\eqref{eq:y2} to characterise the moment at which the non-linear backreaction effects become so large as to end the linear regime and the tachyonic amplification. We will be able to define a univocal criterion that can be applied to the analytical solution as well as to the numerical results, thus allowing a direct comparison.

\subsection{Non-linear regime}\label{sec:lattice}

As quantum fluctuations grow during the tachyonic phase, the previously-neglected quartic self-interaction term in Eq.~\eqref{eq:Yeom} becomes increasingly important, up to the point of completely halting the tachyonic growth of perturbations and marking the beginning of the oscillations of the spectator field. The assumptions at the core of the linear analysis in Section \ref{sec:linear_regime} cease then to be valid. In order to study this highly non-linear regime, we will resort in what follows to fully-fledged classical lattice simulations in $3+1$ dimensions. To this end, we will make use of a modified version of the publicly-available \texttt{$\mathcal{C}osmo\mathcal{L}attice$} code \cite{cosmolattice, Figueroa:2016wxr}, designed to evolve interacting scalar fields in an expanding background. In particular, we modified the default lattice implementation of the Klein-Gordon equation in this numerical tool to account for the Hubble-dependent mass term in Eq.~\eqref{eq:eomchi_scalar}, introducing also related definitions of the energy density and pressure in Eqs.~\eqref{eq:chi_energy_density_pressure_scalar0} and \eqref{eq:chi_energy_density_pressure_scalar}. 

In all our simulations, the lattice parameters are chosen to ensure the stability and reliability of the output. In particular:
\begin{enumerate} 
\item  The comoving grid size $L$ and the number of lattice points per dimension $N$ are selected in such a way that all relevant modes are always well within the associated infrared (IR) and ultraviolet (UV) resolution in momentum space, 
\begin{equation}
\kappa_{\rm IR}=\frac{2\pi}{L}\,,\hspace{15mm} \kappa_{\rm UV}= \frac{\sqrt{3}}{2} N \kappa_{\rm IR}  \,. 
\end{equation} 
In particular, these quantities are set to properly cover the tachyonic band found in Eq.~\eqref{tachband} as well as modes enhanced by subsequent rescattering effects. Therefore, following the results of the linear analysis, we identify the smallest momentum in the tachyonic band with {$\kappa_{\text{IR}} = \mathcal{H}$, while the largest amplified momentum is set to be smaller than the lattice's UV momentum, i.e. $\sqrt{4\nu^2 - 1}\mathcal{H} \ll \kappa_{\text{UV}}$}. The last condition implies a constraint on the minimum number of lattice sites $N>2\sqrt{4\nu^2 - 1}/\sqrt{3}$ which is always fulfilled in our simulations, as much larger lattices are needed to cover the late-time dynamics of the system. Since the mode excitation depends mainly on the non-minimal coupling parameter $\nu$, we set $N=128$ for $\nu<10$, $N=256$ for $10\leq \nu < 25$ and $N=512$ for $\nu\geq 25$. In order to verify that our results do not depend on the lattice size, we perform when possible a few test simulations with increased lattice size.
\item 
The time-step variable is chosen according to the stability criterion $\delta t / \delta x \ll 1/\sqrt{d}$ \cite{cosmolattice}, with $d=3$ the number of spatial dimensions and
    \begin{equation}
        \delta x = \frac{2\pi}{N\,\kappa_{\text{IR}}}\Bigg\vert_{z=0} = \frac{4\pi \, \nu }{N}
    \end{equation}
     the length of the side of a lattice cell. More specifically,  we set $\delta t=0.1$ for $\nu \geq 10$ and $\delta t=0.01$ for $\nu < 10$. 
\end{enumerate}
Given the fixed power-law background expansion in \eqref{eq:paramexpansion}, a symplectic 4th order Velocity-Verlet evolver guarantees stability and precision of the numerical solutions when the conservation of energy cannot be explicitly checked. In agreement with the overall picture developed in the previous section, the initial conditions for our lattice simulations are set as $\chi(0)=\chi'(0)=0$, with fluctuations over this homogeneous background included as Gaussian random fields, as done customarily for systems with short classicalisation times like the one under consideration \cite{Bettoni:2021zhq}. The resulting evolution is therefore deterministic up to a randomly-generated initial seed. A robust (and time-consuming) approach would require to perform several realisations of the same simulation, averaging subsequently over the final output to make sure that the random initial conditions do not influence the dynamics. However, because of the computational resources at our disposal, we choose to keep the base seed constant in all our simulations, making them exactly comparable. Since the system looses memory of the initial conditions soon after the development of the tachyonic instability, this choice does not influence the overall macroscopic evolution, as we have explicitly checked by performing a few simulations with different base seeds. 

\subsection{The timescales of heating}\label{sec:reheating_scalar}

In this section, we present the main results of our extensive parameter scanning in the $(\nu, \lambda)$ plane, obtained by performing repeated 3+1 dimensional classical lattice simulations with $\lambda \in [10^{-6},\, 10^{-1}]$ and ${\nu \in [5,\, 35]}$. The domain of $\nu$ is set to avoid sizeable horizon-reentry effects ($\nu \geq 5$) while ensuring the covering of all relevant modes throughout the evolution ($\nu \leq 35$). As will become clear in what follows, the main timescales in the evolution of the system for $\nu>35$ can be inferred from simulations with smaller non-minimal couplings. The range of $\lambda$ is chosen to ensure perturbativity ($\lambda<1$) while partially covering the typical values appearing in Standard Model settings ($10^{-6}\lesssim \lambda \lesssim 10^{-1}$), namely those of the Higgs self-coupling at energies around $10^{10}-10^{16}$ GeV  \cite{ATLAS:2019guf, CMS:2019esx, Bezrukov:2014ina}. In principle, the requirements of energetic subdominance of the spectator field and of sub-Planckian contribution to the reduced Planck mass set boundaries on the parameter space as well. However, being these boundaries dependent on the scale of kination $H_{\rm kin}$, we can proceed independently from its exact value and work with dimensionless quantities such as $\chi(z)/H_{\rm kin}$ and $\rho(z)/H^4_{\rm kin}$, as we did in the previous section through $\chi_{\rm kin}$. The discussion on the correction to the Planck mass and to the energetic subdominance of the spectator field will be made \emph{a posteriori}.

\begin{figure}[t]
    \centering
    \includegraphics[width=0.8\textwidth]{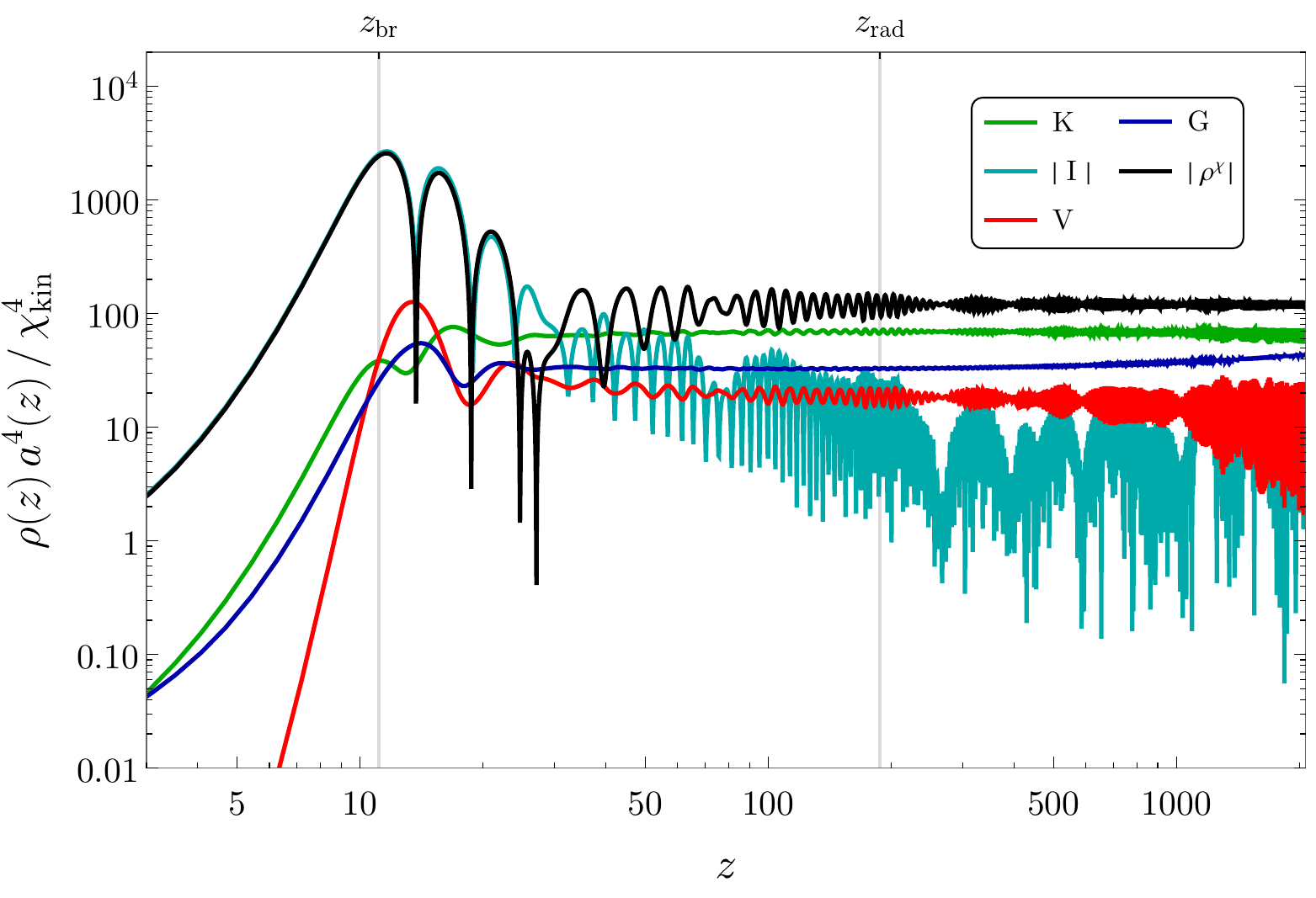}
    \caption{
Evolution of the volume-averaged kinetic ($K$), gradient ($G$), potential ($V$) and interaction ($I$) contributions in \eqref{eq:energy_components} to the total energy density $\rho^{\chi}$ for exemplary values $\nu=10 $ and $\lambda =10^{-4}$. Each energy component is multiplied by $a^4$ to highlight asymptotic radiation-like behaviours. In order to facilitate the comparison, we display the absolute values of the interaction and the total-energy terms, since they are not positive definite at all times. The timescales identifying the backreaction time $z_{\rm br}$ and the virialisation time $z_{\rm rad}$ for this specific case are indicated at the top of the image.}
    \label{fig:single_scalar_energies}
\end{figure}

A typical output of our simulations is presented in Fig.~\ref{fig:single_scalar_energies}, where we display the evolution of the spectator-field energy density, separated in its kinetic ($K$), gradient ($G$), potential ($V$) and interaction ($I$) contributions,
\begin{equation}
\begin{aligned}
  &  K=\frac{1}{2} \chi'^2 \, , \hspace{5mm} G=\frac{1}{2a} (\nabla\chi)^2 \,, \hspace{5mm} V=\frac{\lambda}{4} \chi^4 \, , \hspace{5mm}  I=3\xi\mathcal{H}\chi^2 +6\xi\mathcal{H}\chi\chi' - \frac{\xi}{a^2}\nabla^2\chi^2 \, .
    \label{eq:energy_components}
\end{aligned}
\end{equation}
With these definitions, the interaction term $I$ contains a total-divergence piece that averages to zero once integrated over the full lattice volume with periodic boundary conditions \cite{Repond:2016sol, Figueroa:2021iwm}. Note that both the interaction term $I$ and the total energy $\rho^{\chi}$ are displayed as absolute values in Fig.~\ref{fig:single_scalar_energies}, since the non-minimal coupling (tachyonic) contribution is not positive definite. However, the sum of the inflaton and spectator fields energy densities is always positive definite \cite{Ford:1987de,Ford:2000xg,Bekenstein:1975ww,Flanagan:1996gw}. Note also that in Fig.~\ref{fig:single_scalar_energies} each term has been multiplied by a $a^{4}$ factor, in order to highlight the radiation-like evolution of the kinetic and gradient terms at late times.

The presence of the non-minimal coupling to gravity leads to a rapid growth of fluctuations that comes to an end when the non-linearities associated to the quartic self-interaction term become relevant. The \emph{backreaction time} $z_{\text{br}}$ can be identified with the first time at which the second time-derivative of the spectator field vanishes,  $ Y''(z)=0$, which is itself related to the time at which the effective frequency of the field's oscillations vanishes,
    \begin{equation}
     \left( -\nabla^2 Y + \lambda Y^3 \right) \Big|_{z_{\text{br}}} =  M^2(z)Y  \Big|_{z_{\text{br}}} \,. \label{zbredef}
    \end{equation}
Contrary to other choices in the literature, this definition can be conveniently applied to both analytical and numerical results, enabling their direct comparison.~\footnote{ One could consider other definitions of the backreaction timescale based on the the amplitude of the scalar field's oscillations, on the evolution of the effective mass or on the different contributions to the total energy density (see for example Refs.~\cite{Figueroa:2015rqa, Kofman:1997yn}). We have verified that, due to the fast growth of tachyonic perturbations, such alternative definitions translate into negligible differences within our numerical results.} In particular, due to the stochastic nature of the lattice approach, we choose to consider the lattice-averaged amplitude of the fluctuations as a good indicator of the overall dynamical evolution, comparing the time at which the root-mean-squared $\langle Y(z)' \rangle_{\text{rms}}$ reaches its maximum value to the analytical definition of $z_{\text{br}}$ following from
    \begin{equation}
      \kappa_{\ast}^2  + \lambda  Y^2(z_{\text{br}})  =  M^2(z_{\text{br}})\,,
      \label{eq:analyticzvev_scalar}
    \end{equation}
    with $\kappa_{\ast}=2\sqrt{2\nu +1}\mathcal{H}$ the typical momentum scale entering the analytical two-point correlation function in Eq.~\eqref{eq:size0}.

 \begin{figure}[t]
        \centering
        \includegraphics[width=0.8\textwidth]{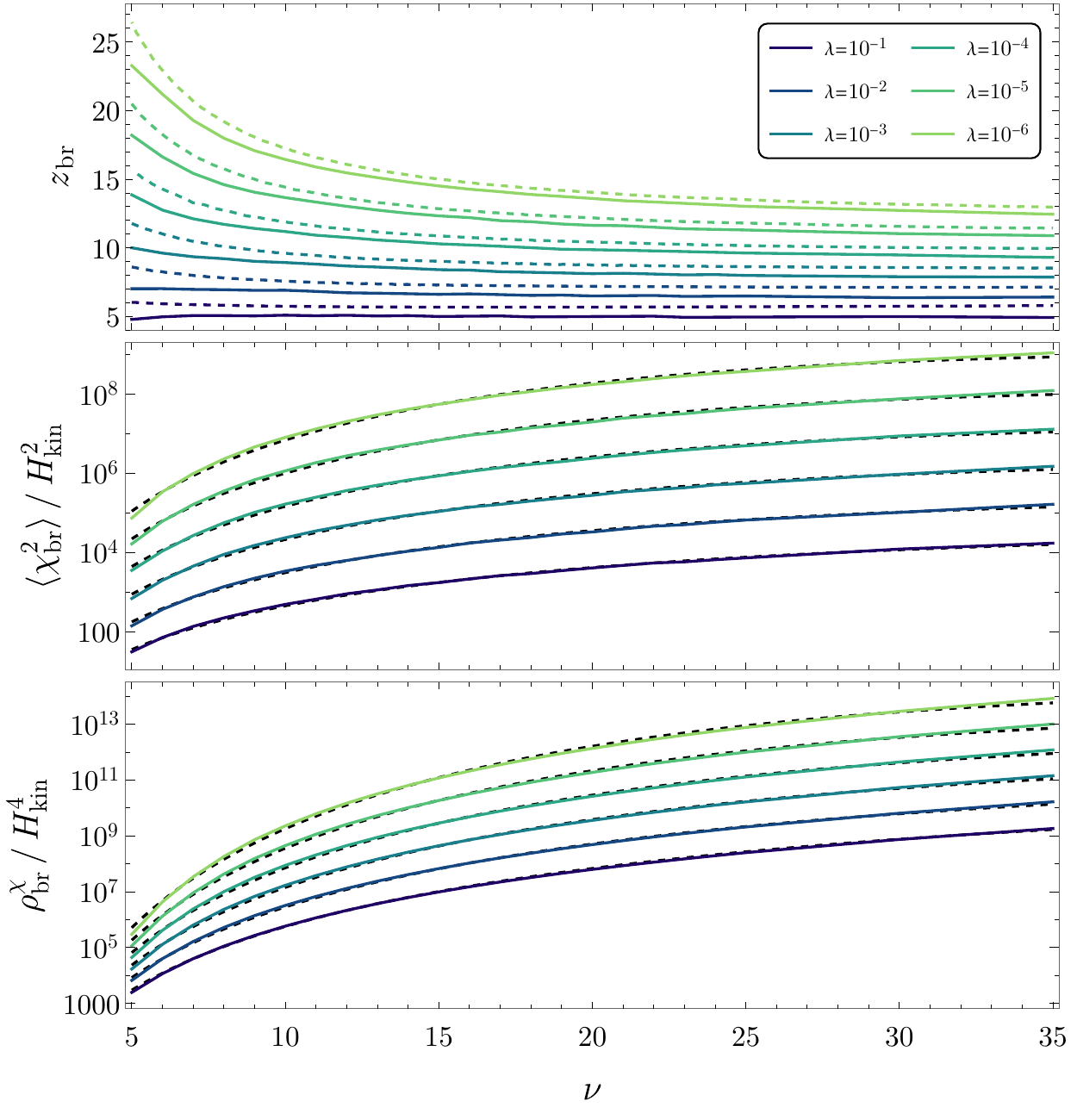}
        \caption{Parametric dependence of the backreaction timescale $z_{\text{br}}$ (upper panel), of the amplitude of the scalar field fluctuations $\langle \chi^2_{\text{br}}\rangle_{\text{rms}}$ (middle panel) and of the total energy at backreaction $\rho^{\chi}_{\text{br}}$ (lower panel) on the model parameters $\nu$ and $\lambda$, as indicated in the figure. In the upper panel, dashed coloured lines correspond to the analytical curve obtained from Eq.~\eqref{eq:analyticzvev_scalar}. In the middle and lower panels, solid lines are derived from the numerical output of lattice simulations while the black dashed lines correspond to the fitting formulas \eqref{eq:fit_Y2_br} and \eqref{eq:fit_rho_br}.}
    \label{fig:backreaction_total}
\end{figure}

The behaviour of the backreaction time $z_{\text{br}}$ following from the analytical expression \eqref{eq:analyticzvev_scalar} is displayed in the upper panel of Fig.~\ref{fig:backreaction_total} as a function of the model parameters in the scan. We observe that, due to the fast-growth of tachyonic modes upon symmetry breaking, this quantity is roughly comparable with the duration of the first semi-oscillation, ${z_{\text{br}}\approx \mathcal{O}(10)}$. The explosive enhancement of fluctuations reaches, however, a saturation point at large $\nu$, with different lines approaching an almost constant  value in one-to-one correspondence with the choice of $\lambda$. Consequently, for non-minimal couplings $\nu \gtrsim 20$, the difference in backreaction times at constant $\lambda$ becomes completely negligible. We also recall here that our analytical estimate of the backreaction scale relies on the large-$\nu$ limit, a factor that can contribute to the small discrepancies in the interval $5 \leq \nu \leq 10$. 

From the data points in Fig.~\ref{fig:backreaction_total} we can additionally derive fitting formulas for the amplitude of the spectator field at backreaction time 
\begin{equation}
    \langle \chi^2_{\rm br} (\lambda, \nu)\rangle_{\text{rms}} = 4 H^2_{\rm kin} \, \exp\left(\alpha_1 + \alpha_2 \nu +{\alpha_3} \ln \nu \right)\,  \,, 
\label{eq:fit_Y2_br}
\end{equation} 
and the total energy of the spectator field at that moment, 
\begin{equation}
    \rho^{\chi}_{\text{br}}(\lambda, \nu) = 16 H^4_{\rm kin} \, \exp\left(\beta_1 + \beta_2 \,\nu+ {\beta_3} \ln \nu \right)  \,. 
\label{eq:fit_rho_br}
\end{equation}    
The dependence on the quartic self-coupling parameter $\lambda$ is encoded in the coefficients
\begin{equation}
    \alpha_1(\lambda) = -4.92 + 0.74 \, n  \,,\hspace{7mm}
    \alpha_2(\lambda) = -0.04 - 0.02 \, n \,, \hspace{7mm}
    \alpha_3(\lambda) = 3.54 + 0.61 \, n \,, \label{coeff_Y2_br}
\end{equation}
and
\begin{equation}
    \beta_1(\lambda) = -7.03 - 0.56 \, n  \,,\hspace{7mm}
    \beta_2(\lambda) = -0.06 - 0.04 \, n \,, \hspace{7mm}
    \beta_3(\lambda) = 7.15 + 1.10 \, n \,, \label{coeff_rho_br}
\end{equation}
which are all simple first-order polynomials in $n=-\log(\lambda)$. These formulas are displayed as black dashed lines in the middle and lower panels of Fig.~\ref{fig:backreaction_total}. Note that the mild dependence on the quartic self-coupling allows us to obtain a reliable fit on a logarithmic interval for $\lambda$ that spans six orders of magnitude.

 \begin{figure}[t]
        \centering
        \includegraphics[width=0.8\textwidth]{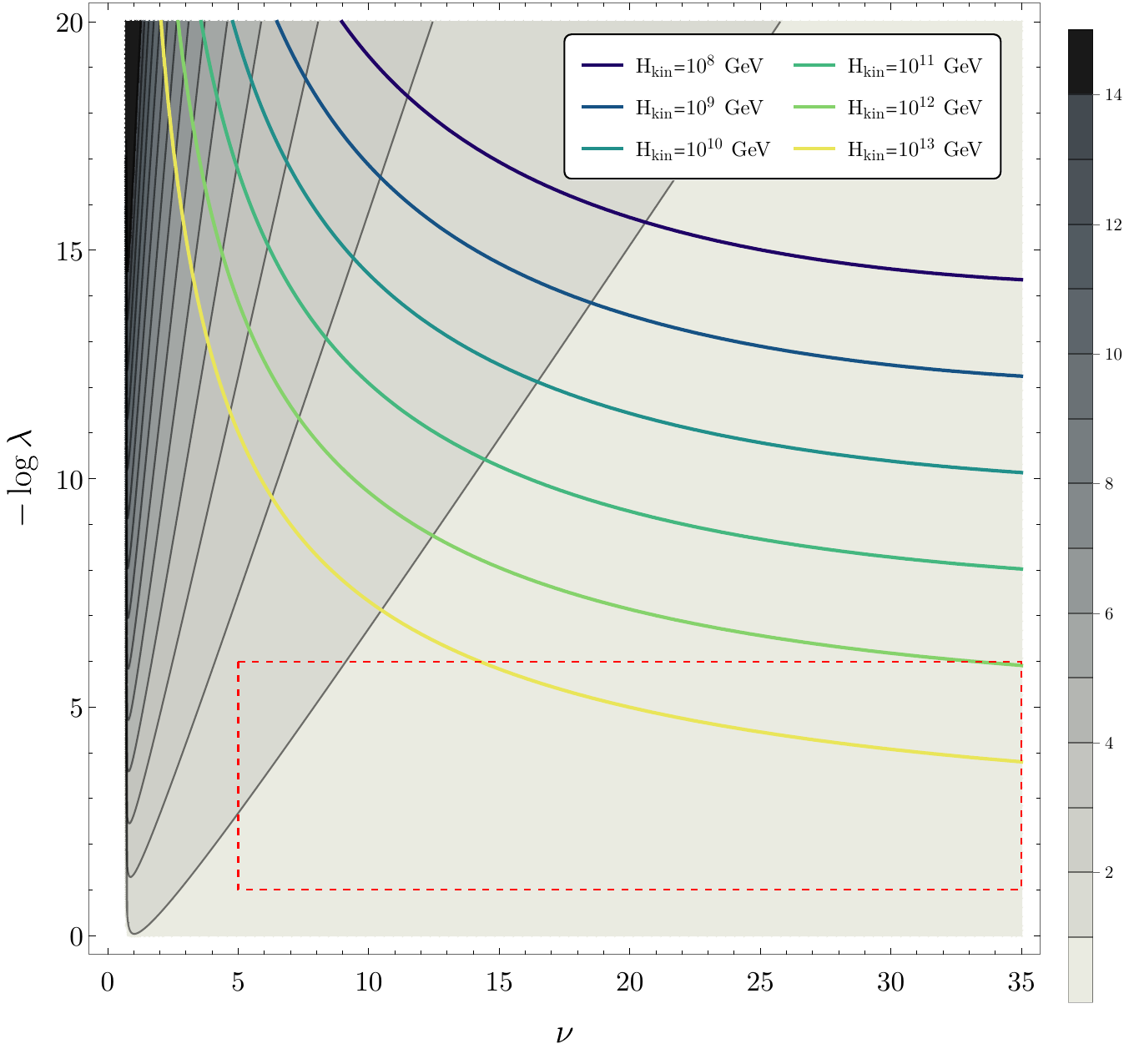}
        \caption{Correction to the squared Planck mass given by the peak mode enhancement at backreaction $\langle \chi^2_{\rm br}\rangle_{\text{rms}}$ as a function of the model parameters. The coloured lines set the contours for a $10\%$ correction at different values of the Hubble scale at the onset of kination. The grey shading indicates the duration of the tachyonic phase in $e$-folds $N_{\rm br}=1/2 \ln(1+z_{\rm br}/\nu)$. The dashed red rectangle delimits the range explored in our extensive scanning of the parameter space.}
    \label{fig:corr_planck_mass}
\end{figure}

As mentioned previously, the self-consistency of our procedure requires the explosive tachyonic enhancement of spectator field fluctuations to fulfil the condition on negligible contributions to the reduced Planck mass. The parametric formula \eqref{eq:fit_Y2_br} allows us to estimate the peak amplification of the scalar field and with it the effective change in $M^2_P$ at the level of the model's Lagrangian. Figure~\ref{fig:corr_planck_mass} shows with solid lines the $10\%$ threshold for different choices of the energy scale at the beginning of kination. Given the self-similar evolution of the lines representing $\langle \chi^2_{\rm br}\rangle_{\text{rms}}$ in Fig.~\ref{fig:backreaction_total}, we can extrapolate the maximal amplitude even for $\lambda < 10^{-6}$, thanks to the parametric dependence in \eqref{eq:fit_Y2_br}. The red rectangle indicates the actual region explored in our scanning, which confirms that our setup is fully consistent with a subdominant scalar field for $H_{\rm kin}\leq 10^{12} \, \rm GeV$.

For $z>z_{\text{br}}$, the system enters a highly-oscillatory regime driven initially by the Hubble-induced contributions (cf.~Fig.~\ref{fig:single_scalar_energies}). The impact of these time-dependent terms becomes, however, completely negligible at later times, where the steady flow of energy from the IR to the UV part of the spectrum effectively virializes the system \cite{Bettoni:2021zhq}, thus approaching an asymptotic energy configuration with $\langle K \rangle \approx \langle G \rangle + 2\langle V \rangle$. 
This condition provides a natural lattice definition for the \emph{radiation time} $z_{\text{rad}}$ at which the spectator field starts behaving as a radiation fluid, namely
    \begin{equation}
        \biggl< \frac{\langle K(z_{\text{rad}}) \rangle }{\langle G(z_{\text{rad}})\rangle + 2\langle V(z_{\text{rad}})\rangle} \biggr>_{t} = 1 \pm 0.01 \, ,
    \label{eq:vir_criterium_lattice}
    \end{equation}
with the subscript $t$ denoting a time-average over several oscillations of the scalar field. Note that this indirect definition of $z_{\rm rad}$ holds as long as the contribution of the non-minimal coupling terms is negligible, coinciding in that limit with the usual condition $w_{\chi}(z_{\text{rad}})=1/3$, as can be easily inferred from Eqs.~\eqref{eq:chi_energy_density_pressure_scalar0} and \eqref{eq:chi_energy_density_pressure_scalar}.  Fig.~\ref{fig:single_scalar_energies} shows that the condition \eqref{eq:vir_criterium_lattice} is first fulfilled at $z_{\text{rad}} \approx \mathcal{O}(100)$, when the energy stored in the non-minimal interaction is at least one order of magnitude smaller than the kinetic, gradient and potential counterparts. 

 \begin{figure}[t]
        \centering
        \includegraphics[width=0.8\textwidth]{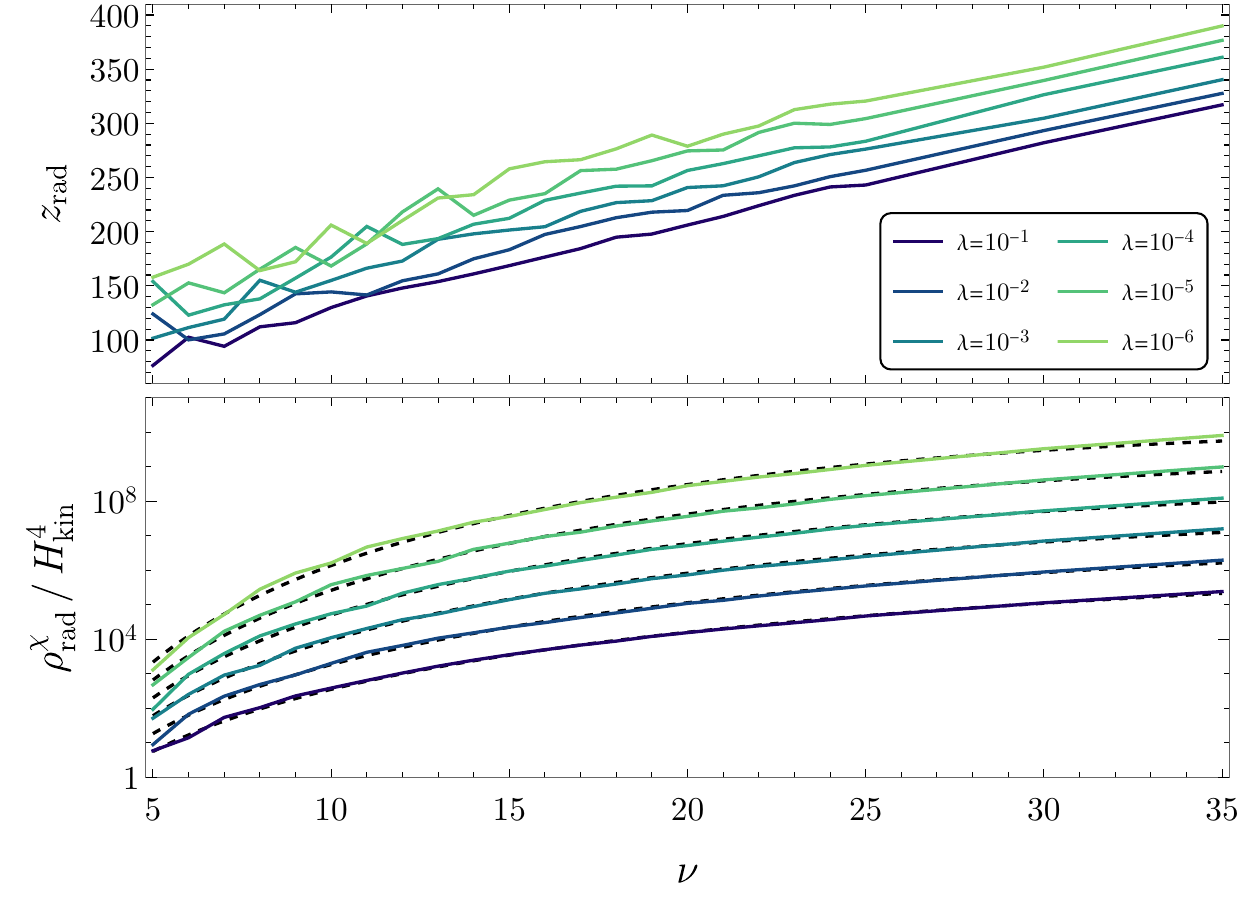}
        \caption{Parametric dependence of the backreaction timescale $z_{\text{rad}}$ (upper panel), the total energy at completion of virialisation $\rho^{\chi}_{\text{rad}}$ (lower panel) on the model parameters $\nu$ and $\lambda$, as indicated in the figure. Solid coloured lines correspond to the numerical results obtained from the full set of lattice simulations. In the lower panel, black dashed lines correspond to the fitting formula \eqref{eq:fit_rho_rad}}.
    \label{fig:radiation_total}
\end{figure}

The specific dependence of the radiation time $z_{\text{rad}}$ on the model parameters in the scan is displayed in the top panel of Fig.~\ref{fig:radiation_total}. In deriving these results we have checked that the contribution of the non-minimal coupling 
becomes subleading as the original symmetry of the theory is restored and the Hubble function decreases with time. This translates into the condition $I < 0.1 \times K, \, G, \, V$ which is satisfied by all the data points in the figure. For small values of $\nu$, the virialisation process is completed soon after the non-linear regime sets in. At that stage, the spectator field is still undergoing large oscillations, making the precise determination of $z_{\rm rad}$  more difficult and explaining the irregular features at $\nu<15$. We note also that, due to a stronger rescattering of modes, $z_{\text{rad}}$ is generally shorter for higher values of $\lambda$. 

The above results allow us to derive fitting formulas for the timescale $z_{\text{rad}}$ as function of the model parameters. For all values of $\lambda$, the behaviour of $z_{\text{rad}}$ deduced from Fig.~\ref{fig:radiation_total} is linear in $\nu$, namely
\begin{equation}
    z_{\text{rad}}(\lambda, \nu) = \gamma_1 + \gamma_2 \, \nu \,, 
\label{eq:fit_z_rad}
\end{equation}   
 with
\begin{equation}
    \gamma_1(\lambda) = 33.63 + 15.02 \, n - 0.22 \, n^2 \,,\hspace{10mm}
    \gamma_2(\lambda) = 7.91 - 0.01 \, n + 0.02 \, n^2 \,, \label{c12rad}
\end{equation}
second-order polynomials in $n=-\log (\lambda)$. We can additionally derive a fitting formula for the total energy of the spectator field at $z_{\text{rad}}$, namely 
\begin{equation}
    \rho^{\chi}_{\text{rad}}(\lambda, \nu) = 16 H^4_{\rm kin} \, \exp\left({\delta_1 + \delta_2 \, \nu} +{\delta_3}\ln \nu\right) \,, 
\label{eq:fit_rho_rad}
\end{equation}  
with
\begin{equation}
    \delta_1(\lambda) = -11.10 - 0.06 \, n  \,,\hspace{5mm}
    \delta_2(\lambda) = -0.04 - 0.03 \, n \,, \hspace{5mm}
    \delta_3(\lambda) = 5.62 + 0.87 \, n \,. \label{coeff_rho_rad}
\end{equation}
This is displayed in the bottom panel of Fig.~\ref{fig:radiation_total} as black dashed lines. By comparing the bottom panels of Fig.~\ref{fig:backreaction_total} and Fig.~\ref{fig:radiation_total}, we notice that the total energy available at the end of the virialisation process is roughly a constant fraction of the energy produced during the tachyonic phase. In particular, the fitting formulas \eqref{eq:fit_rho_br} and \eqref{eq:fit_rho_rad} tell us that ${\rho^{\chi}_{\text{rad}}(\lambda, \nu) \sim 10^{-4} \times \rho^{\chi}_{\text{br}}(\lambda, \nu)}$.

 \begin{figure}[t]
        \centering
        \includegraphics[width=0.75\textwidth]{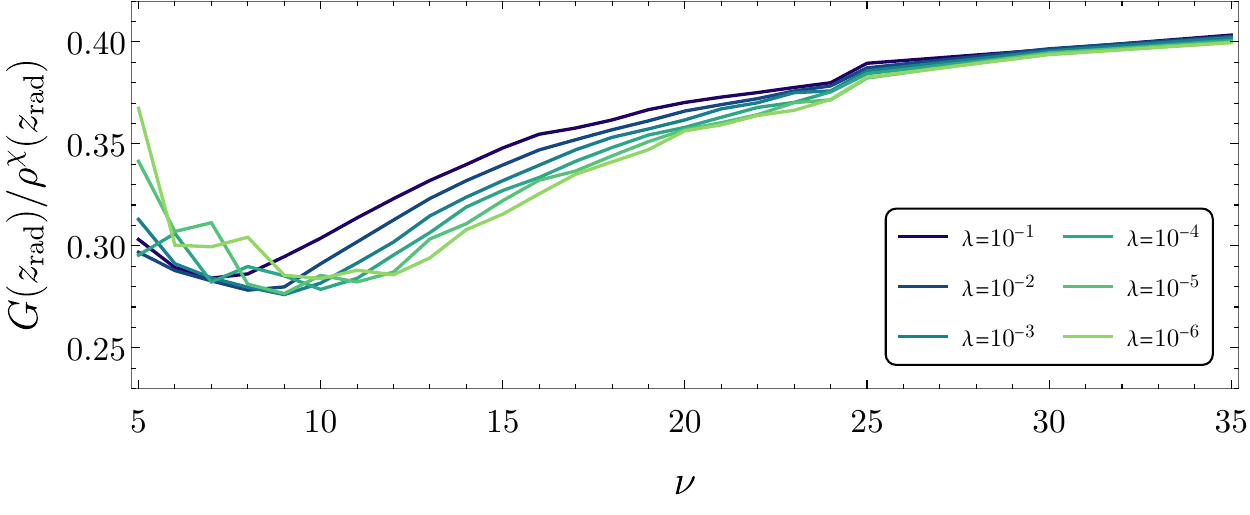}
        \caption{Ratio of the energy density stored in gradients and the total energy density of the spectator field at $z_{\rm rad}$, as a function of the model parameters $\nu$ and $\lambda$. }
    \label{fig:grad_kin_ratio}
\end{figure} 

At this stage, it is instructive to compare our finding with the results of an homogeneous treatment, where the system dynamics gives rise to an effective equation-of-state parameter $w_\chi=1/3$ soon after the onset of the oscillatory regime \cite{Opferkuch:2019zbd, Bettoni:2021zhq}. As apparent in Fig.~\ref{fig:single_scalar_energies}, the presence of spatial inhomogeneities is certainly not a negligible effect, with the energy density into gradients accounting to a sizeable fraction of the total energy density and generically exceeding the potential energy contribution. This is confirmed in Fig.~\ref{fig:grad_kin_ratio}, where we display the gradient-to-total energy ratio $G/\rho^{\chi}$ computed at the radiation time $z_{\rm rad}$ for all the model parameters in the scan. Since the initial tachyonic production is more effective at sourcing spatial inhomogeneities, this ratio is generically larger for larger values of $\nu$, leading to gradient contributions of up to 40$\%$ of the total energy density. The kink feature visible at $\nu=25$ is a mild numerical effect due to the change of lattice size, which slightly modifies the relative importance of gradients. 

\section{Heating efficiency and temperature}\label{sec:radtemperature}

\begin{figure}[t]
    \centering
    \includegraphics[width=0.8\textwidth]{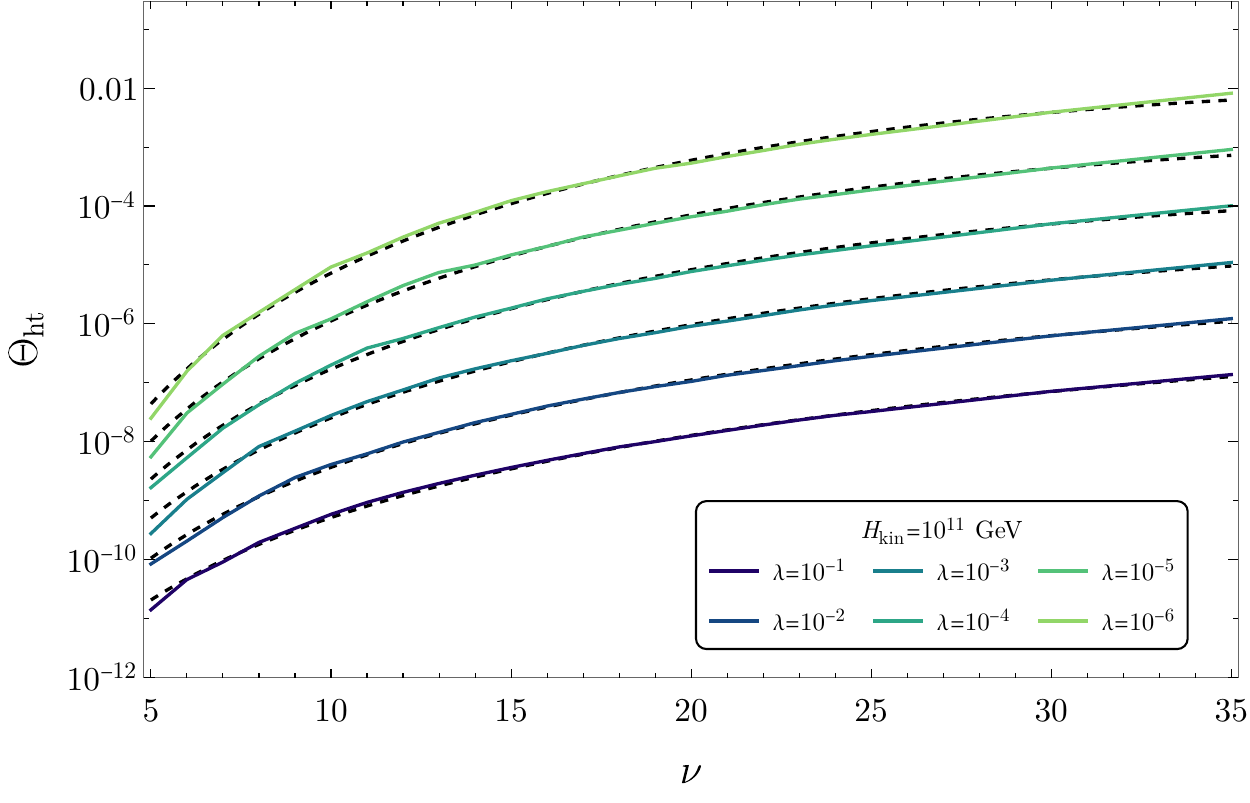}
    \caption{Parametric dependence of the heating efficiency $\Theta_{\text{ht}}$ on the model parameters $\lambda$ and $\nu$ for a fiducial energy scale of kination $H_{\text{kin}} = 10^{11}$ GeV. Solid lines correspond to the values computed from the simulations scanning of the parameter space, while the dashed lines correspond to the fitting formula \eqref{eq:fittingTheta}.}
    \label{fig:efficiency_temp_scalar}  
\end{figure}

After reaching a relativistic equation of state at $z_{\rm rad}$, the spectator field's energy density evolves proportionally to $a^{-4}$, with the associated power spectrum approaching slowly a thermal distribution \cite{Micha:2002ey, Micha:2003ws, Micha:2004bv}. From there on, the evolution of the Universe's energy budget can be studied analytically via the so-called \emph{heating efficiency} introduced in Ref.~\cite{Rubio:2017gty}. The most straightforward definition of this quantity is based on the ratio of the energy density of the spectator field \emph{at the onset of kination} to that of the inflaton component, ${\Theta_{\text{ht}} = \rho^{\psi}(a_{\rm kin}) / \rho^{\phi}(a_{\rm kin})}$, which is valid in case of instantaneous production of relativistic daughter fields $\psi$ at the end of inflation \cite{Bettoni:2021qfs, Rubio:2017gty}. However, in the present model the tachyonic production of $\chi$ particles is completed in about one $e$-fold after the onset of kination, taking another $e$-fold for the spectator field to reach a radiation-like equation of state. 
Therefore, it is more convenient to define the heating efficiency \emph{at the onset of the radiation-like behaviour} namely 
\begin{equation} \label{eq:heating_eff_def}
    \Theta_{\text{ht}} \equiv \frac{\rho^{\chi}(a_{\rm rad})}{\rho^{\phi}(a_{\rm rad})} \,, 
\end{equation}
with $\rho^{\phi}(a)=3 M^2_{P}H^2_{\rm kin}(a/a_{\rm kin})^{-6}$ the energy density of the inflaton field in the kinetic dominated regime. Note that, since $\Theta_{\text{ht}}|_{\rm rad}=\Theta_{\text{ht}}|_{\rm kin} \, (a_{\rm rad}/a_{\rm kin})^2$, it is always possible to switch between the two alternative definitions. By inserting the parametric formulas \eqref{eq:fit_z_rad} and \eqref{eq:fit_rho_rad} in Eq.~\eqref{eq:heating_eff_def} we gain also a parametric expression for $\Theta_{\rm ht}$,
\begin{equation} \label{eq:fittingTheta}
    \Theta_{\text{ht}}(\lambda, \nu) = \frac{16}{3}\left(\frac{H_{\rm kin}}{M_{P}}\right)^2  \left(1+\frac{\gamma_1 + \gamma_2 \, \nu}{\nu}\right)^3 \exp\left({\delta_1 + \delta_2 \, \nu} +\delta_3 \ln \nu \right)  \,,
\end{equation}
showing explicitly its dependence on the model parameters. As displayed in Fig.~\ref{fig:efficiency_temp_scalar} for the specific case of $H_{\rm kin}=10^{11} \; \rm GeV$, the computed values fulfil the big bang nucleosynthesis bound $\Theta_{\text{ht}} \gtrsim 10^{-16}$ derived in Ref.~\cite{Bettoni:2021qfs}. Higher values of this quantity correspond to longer and stronger tachyonic phases following from large-$\nu$ and small-$\lambda$ regimes, as demonstrated by the tight correlation between $\Theta_{\text{ht}}(\lambda, \nu)$, $\rho_{\text{rad}}(\lambda, \nu)$ and $\rho_{\text{br}}(\lambda, \nu)$ in Figs.~\ref{fig:backreaction_total}, \ref{fig:radiation_total} and \ref{fig:efficiency_temp_scalar}. 

The expression for the heating efficiency in Eq.~\eqref{eq:fittingTheta}, together with Eqs.~\eqref{eq:fit_Y2_br}, \eqref{eq:fit_rho_br}, \eqref{eq:fit_z_rad} and \eqref{eq:fit_rho_rad}, constitutes the central result of this work. The timescales corresponding to backreaction time and to radiation time mark the interval where the complicated non-linear dynamics has to be solved with numerical simulations. The initial tachyonic phase is well captured by the analytical solution to the mode equation in \eqref{eq:solY}. The parametric formulas for $\rho^{\chi}_{\rm rad}$ and $z_{\rm rad}$ summarise the macroscopic characteristics of the complicated self-interacting system that we have numerically simulated in more than one hundred runs. The quality of the fitting procedure has been checked for all of our fitting formulas via the $R^2$ method, yielding always $R^2>0.99$. Given the simplicity and reliability of our results, the parametric formulas deliver a ready-to-use description of the main phases of heating, eliminating the need for further simulations over (and beyond) the parameter space we have explored. 

Although the spectator scalar field achieves a radiation-like equation of state at $z_{\rm rad}$, a fully thermalised state is not reached until much later, once the non-linear rescattering processes has redistributed the available energy-density within a large range of modes \cite{Micha:2002ey, Micha:2003ws, Micha:2004bv,Bettoni:2021zhq}. Nonetheless, it is convenient to define an \emph{instantaneous temperature} scale, 
\begin{equation}
T_{\rm ht} =\left(\frac{30\,\rho^{\chi}_{\rm ht}}{\pi^2 g_*^{\rm ht}}\right)^{1/4} \,,   
\end{equation}
with $g_*^{\rm ht}=106.75$ the Standard Model number of relativistic degrees of freedom at energies above ${\cal O}(100)$ GeV and $\rho^{\chi}_{\rm ht}=\rho^{\chi}(z_{\rm ht})=\rho^{\phi}(z_{\rm ht})$ the total energy-density of the scalar field at the end of the heating phase. This temperature should be understood as a mere estimate of the typical energy of a particle at the onset on radiation domination, coinciding with the most common concept of \textit{(re)heating temperature} in the fast thermalisation limit. Independently of its thermal character, this scale plays an important role in the production of dark matter relics via UV-freeze-in \cite{Bernal:2020bfj,Bernal:2020qyu} and many mechanisms aiming to explain the observed matter-antimatter asymmetry of the Universe \cite{Canetti:2012zc}. Assuming that the overall equation of state of the Universe is $w\approx 1$ until the end of heating, we obtain 
\begin{equation} \label{eq:reheating_temp}
    T_{\text{ht}} \simeq 2.7 \times 10^{8} \,\text{GeV} \left(1+\frac{z_{\rm rad}}{\nu}\right)^{-3/4} \, \left( \frac{\Theta_{\text{ht}}}{10^{-8}}\right)^{3/4}  \left( \frac{H_{\rm kin}}{10^{11} \, \text{GeV}}\right)^{1/2}  \,,
\end{equation}
with the heating efficiency $\Theta_{\text{ht}}$ containing an implicit dependence on the scale of kination, cf.~Eq.~\eqref{eq:fittingTheta}. As expected, due to their long-lasting tachyonic instabilities, simulations with smaller values of $\lambda$  are characterised by a significantly higher heating temperature, up to ${T_{\text{ht}} \sim 10^{12} \, \rm GeV}$ for $H_{\rm kin}=10^{11} \, \rm GeV$. However, even the lowest temperature in our simulations, ${T_{\text{ht}} \sim  10^{5} \, \rm GeV}$, falls way above the Planck and big bang nucleosynthesis bounds on the heating temperature, $T_{\text{ht}} \geq 5 \, \rm MeV$ \cite{deSalas:2015glj, Hasegawa:2019jsa}.

The parametric formulas for $\rho^{\chi}_{\rm rad}$ and $z_{\rm rad}$ allow also for the computation of the minimal number of $e$-folds of inflation $N$ needed to solve the flatness and horizon problems in scenarios where the Hubble-induced phase transition is the only particle production channel. As computed in Ref.~\cite{Rubio:2017gty}, this is given by
\begin{align}\label{eq:nefolds_inflation}
    N  = - \ln &\left( \frac{k_{\rm hc}}{a_0 H_0}\right) + \ln \left( \frac{H_{\rm hc}}{H_0}\right) +\frac14 \ln \left( \frac{\rho_{\rm mat}}{\rho_{\rm hc}}\right) \nonumber \\ 
    &+ \ln \left( \frac{a_{\rm mat}} {a_0}\right) - \frac{1}{2} \ln \left( \frac{a_{\rm kin}}{a_{\rm rad}}\right) + \frac{1}{4}\ln \left( \frac{\rho_{\rm hc}}{\rho_{\rm kin}}\right) \,,
\end{align}
with the subscript \emph{hc} referring to quantities evaluated at the horizon crossing of the pivot scale $k_{\rm hc}=0.002\, \text{Mpc}^{-1}$, \emph{ kin}, \emph{rad}, \emph{mat} denoting respectively quantities evaluated at the beginning of the kination, radiation and matter-domination and $0$ indicating their present-day values. Numerically this corresponds to \cite{Rubio:2017gty}
\begin{equation} \label{eq:nefolds_inflation_2}
 N \simeq 63.3 + \frac14 \ln \left( \frac{r}{10^{-3}} \right) - \frac14 \ln \left( \frac{\Theta_{\text{ht}}}{10^{-8}} \right)\, + \frac14 \ln \left(1 + \frac{z_{\rm rad}}{\nu}\right)\,,
\end{equation}
where the parametric formulas \eqref{eq:fit_z_rad} and \eqref{eq:fittingTheta} can be directly included.

\subsection{Impact on inflationary observables}\label{sec:observables}
 
The heating process summarised in the parametric formulas \eqref{eq:fittingTheta} and \eqref{eq:reheating_temp} plays an important role when computing the total duration of inflation, with the contribution of heating efficiency in Eq.~\eqref{eq:nefolds_inflation_2} accounting for up to $\sim 4$ $e$-folds of expansion history for $H_{\rm kin}= 10^{11}\; \rm GeV$. Given the capability of future Cosmic Microwave Background experiments to detect a tensor-to-scalar ratio at the level of $r={\cal O}(10^{-3}-10^{-4})$ while marginally improving the current constraints on the spectral tilt $n_s$ \cite{SimonsObservatory:2018koc,Sugai:2020pjw,CMB-S4:2020lpa}, it might become eventually possible to set bounds on the physics of heating \cite{Drewes:2019rxn,Drewes:2023bbs}. Having this in mind, let us make use of the lattice-based relations \eqref{eq:nefolds_inflation_2} and \eqref{eq:fittingTheta} to determine the precise value of the inflationary observables in a specific quintessential-inflation scenario involving only gravitational particle production. Among the many inflationary models in the literature, one can consider, for instance, the simple $\alpha$-attractor family, 
\begin{equation}
    \mathcal{L_{\phi}}= -\frac{\partial_{\mu}\phi \partial^{\mu}\phi}{2\left( 1- \frac{\phi^2}{6\alpha M^2_P} \right)^2} - V(\phi) \;, 
\end{equation}
encoding, among others, several incarnations of the Higgs inflation paradigm \cite{Rubio:2018ogq,Karananas:2020qkp} and covering a big portion of the $(n_s, \, r)$ space \cite{Kallosh:2019hzo} currently favoured by the Planck/BICEP2/Keck collaborations \cite{BICEP:2021xfz} using a single parameter $\alpha$. The presence of the pole restricts the field's movement in field space, imposing a bound on its value. By transforming to a canonical field, the pole can be effectively shifted to infinity while stretching the potential for the canonical variable near the pole, resulting in the emergence of plateaus \cite{Galante:2014ifa, Linde:2016uec, Artymowski:2016pjz}.

\begin{figure}[t]
    \centering
    \includegraphics[width=0.8\textwidth]{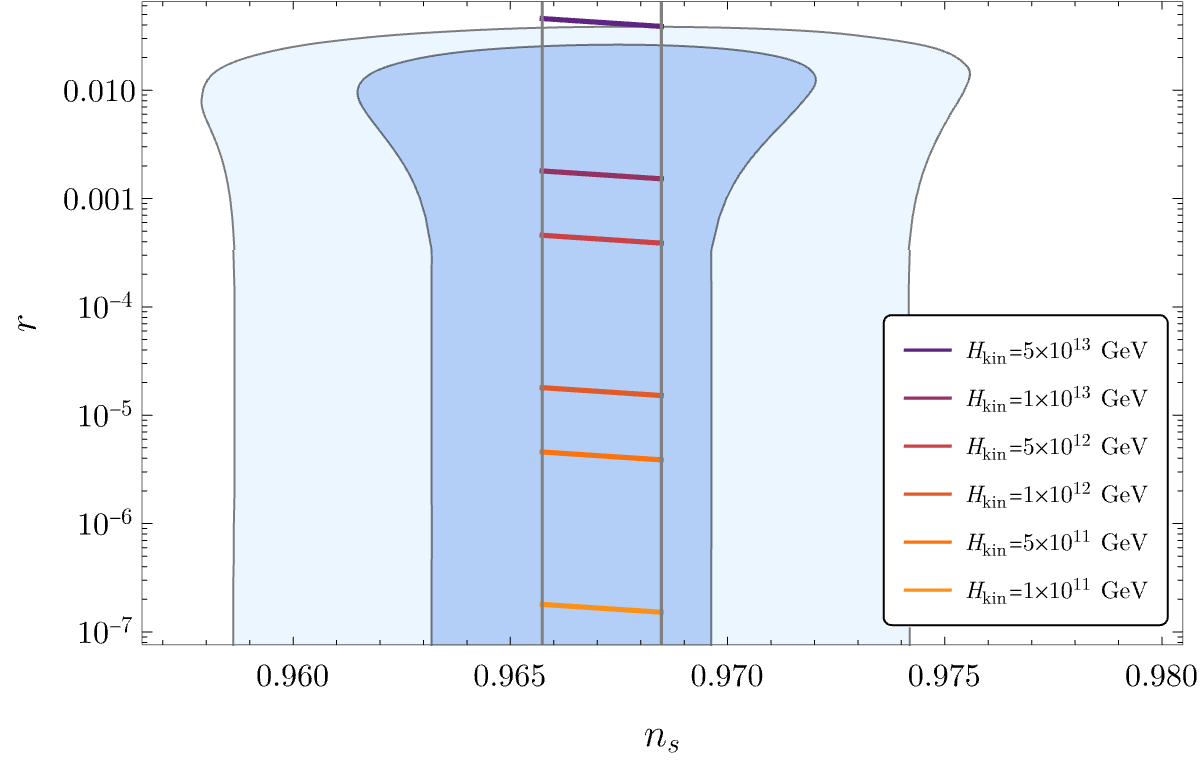}
    \caption{Predicted values of tensor-to-scalar ratio and spectral tilt in an $\alpha$-attractor quintessential-inflation scenario involving only gravitational particle production. Lines of different colours correspond to different values of the $H_{\rm kin}$ parameter, where the end points correspond to the largest (left) and smallest (right) heating efficiencies achieved in our simulations. The vertical grey lines delimit the region allowed by the numerically-obtained heating efficiency. The results are shown in comparison with the constraints from Planck + Bicep/Keck at one and two sigma.}
    \label{fig:planck_bicep_constraints}
\end{figure}

Although initially formulated for oscillatory models, the $\alpha$-attractor idea can be easily adapted to the quintessential inflation paradigm by considering non-symmetric potentials \cite{Akrami:2017cir, Dimopoulos:2017zvq, Garcia-Garcia:2018hlc}. The simplest realisation involves a linear term, adjusted to satisfy the Planck/COBE normalisation constraint, and a small cosmological constant $\Lambda$ responsible for the late-time acceleration of the Universe, namely $
V(\phi)=\gamma \phi + \Lambda$ with $\gamma / (\sqrt{\alpha}M^3_P) \sim 2 \times 10^{-11}$ in order to satisfy the normalisation of the primordial power spectrum  \cite{Akrami:2017cir}. When written in terms of a canonically-normalised field 
$\varphi= \sqrt{6\alpha} M_P\arctanh \phi/(\sqrt{6\alpha}M_P)$,
this potential becomes effectively stretched,
\begin{equation}\label{eq:potentialpha}
    V(\varphi)=\sqrt{6\alpha} \gamma  M_P  \left[ \tanh \left( \frac{\varphi}{\sqrt{6\alpha}M_P} \right) + 1 \right] + \Lambda \,,
\end{equation}
coinciding in the large field limit $\varphi \gg \sqrt{6\alpha} M_P$ with the T-model potential and sharing therefore the very same predictions for the spectral tilt $n_s$, the tensor-to-scalar ratio $r$ and spectral amplitude $\mathcal{A}_s$,
\begin{equation}
    n_s=1-\frac{2}{N} \, , \hspace{10mm} r=\frac{12\alpha}{N^2} \,,  \hspace{10mm} \mathcal{A}_s=\frac{ V_0 N^2}{18 \pi^2 \alpha M^2_P }.
\end{equation}
Combining these expressions with Eqs.~\eqref{eq:fittingTheta} and \eqref{eq:nefolds_inflation_2}, and neglecting the variation of the Hubble rate during inflation, $V_0\simeq 3M^2_P H_{\rm kin}^2$, we can link the inflationary observables to the \textit{heating efficiency} following from our detailed lattice simulations. The result is displayed in Figure~\ref{fig:planck_bicep_constraints}, where we have computed $r$ and $n_s$ assuming specific values for the Hubble scale $H_{\rm kin}$. The vertical grey lines correspond to the predictions for the maximum and minimum number of inflationary $e$-folds ($N_{\rm max}=63.8$, $N_{\rm min}=58.0$) obtained from the minimum and maximum heating efficiency respectively (cf.~Figure~\ref{fig:efficiency_temp_scalar}). The region between the black lines is therefore the full area spanned by the parameter space $(\lambda, \nu)$ of our simulations.

\subsection{Beyond the single field approximation}\label{sec:beyond}
The prototypical case of a $\mathds{Z}_2$-symmetric scalar field considered in the previous section provides us with an extensive understanding of the tachyonic phase (both from the analytical and numerical point of view) and of the timescales of the heating process. This is to be understood as a basic scenario whose phenomenology can be extended to more elaborated field content.

Let us consider, for instance, a scale-free interaction $\frac12 g^2\chi^2\varphi^2$ that respects the global $\mathds{Z}_2$ symmetry of the spectator field, with $\varphi$ some scalar field in a beyond the Standard Model sector and $g^2$ a dimensionless coupling constant. This term, customary in the literature of heating \cite{Kofman:1997yn,Greene:1997fu}, opens up the possibility of depleting the spectator field energy density via perturbative and non-perturbative effects. Given the non-linear nature of the self-scattering interaction and the coupling between fields, let us turn once more to lattice simulations. Due to the fast population of the daughter field's UV modes, a larger lattice is generically needed to cover the evolution of the field's fluctuations, especially in case of large $g$ and prolonged tachyonic phases for $\chi$ (i.e. small $\lambda$). In order to maintain a good coverage of all momenta within a lattice box of $N=256$, we will restrict ourselves to model parameters in the fiducial intervals $g \in [10^{-4}, \, 10^{-3}]$, $\nu \in [10, \, 20]$ and $\lambda \in [10^{-5}, 10^{-1}]$, considering only particular benchmark points due to the longer computational time needed for each simulation.
\begin{figure}[t]
    \centering
    \includegraphics[width=0.8\textwidth]{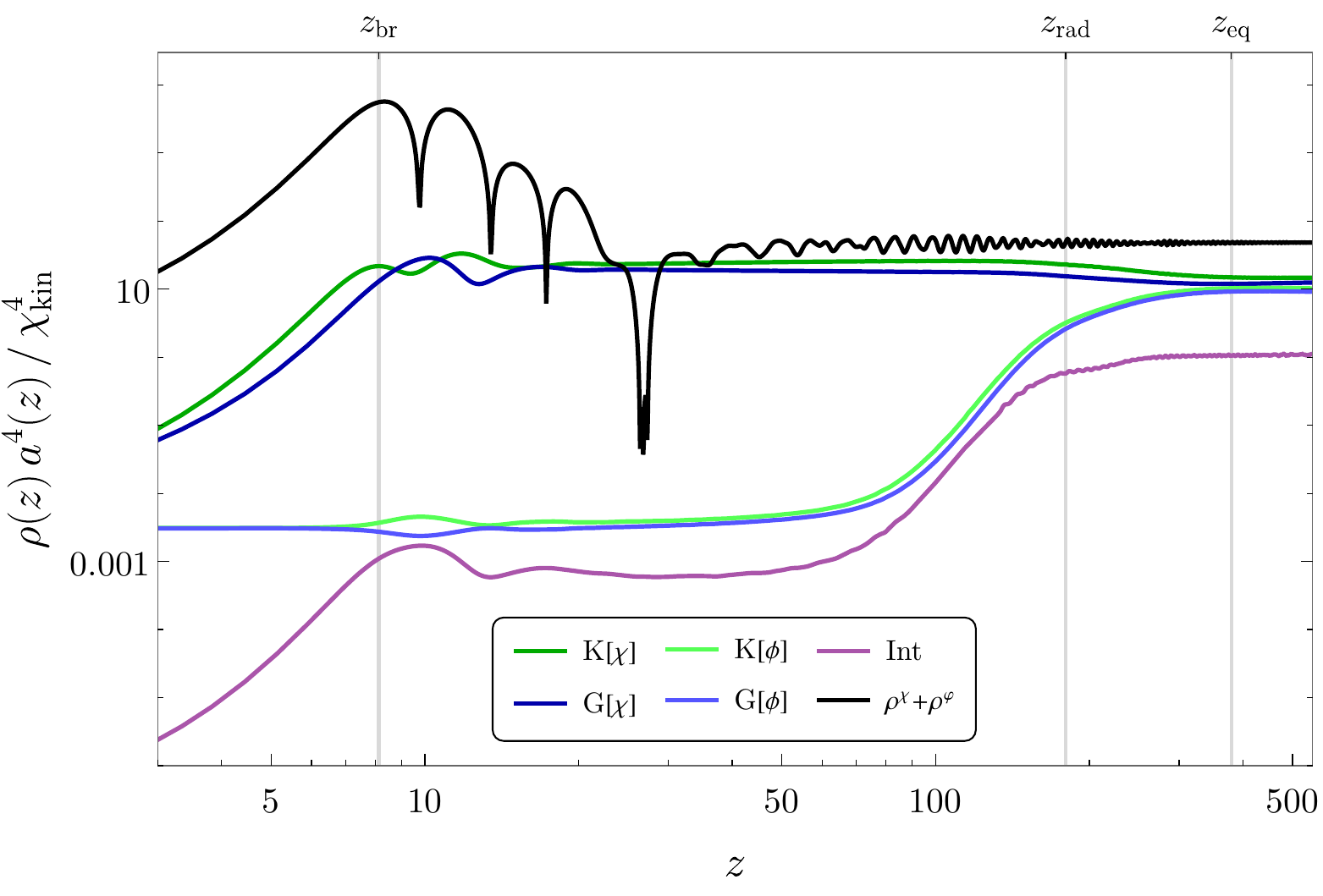}
    \caption{Energy components of the model as defined in \eqref{eq:energy_components}. Each component has been multiplied by $a^4$ in order to highlight the radiation-like behaviour. In this figure, the model parameters have been set to $\nu=20$, $\lambda=10^{-3}$, $g=10^{-3}$.}
    \label{fig.energies_plot_daughter}
\end{figure}

The time-evolution of the different terms contributing to the total energy-density in a typical simulation is displayed in Fig.~\ref{fig.energies_plot_daughter}. As noted already in Section~\ref{sec:reheating_scalar}, the overall evolution of the spectator field is characterised by three main phases: the initial tachyonic amplification of the field's modes, the evolution towards virialisation and the thermalisation of the system. Additionally, we can observe one additional timescale related to the daughter field. The \emph{equipartition} timescale marks the moment when the total energy density is evenly distributed between the two scalar fields. A numerical computation of this timescale is a challenging task, since it can require a very long time to be reached, depending on the production mechanisms and on their efficiency. We simply define it through the condition $\rho^{\chi}(z_{\text{eq}})=\frac{1}{2}\rho^{\rm tot}(z_{\text{eq}})$ for those simulations in which equipartition is reached within their total duration. 

\begin{figure}[tb]
    \centering
    \includegraphics[width=0.9\textwidth]{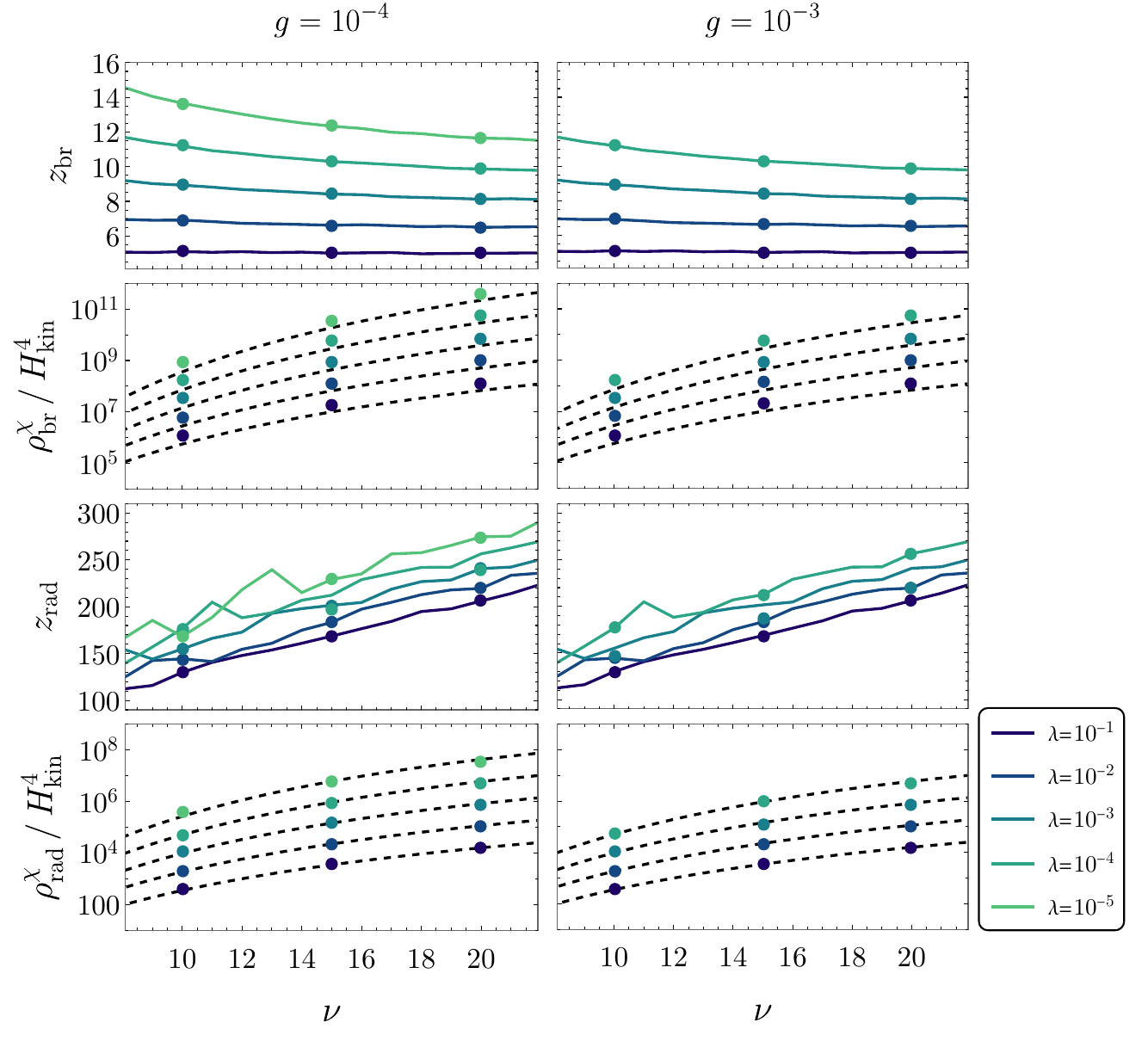}
    \caption{Dependence of the timescales $z_{\text{br}}$ (first panels), $z_{\text{rad}}$ (third panels) and of the energy densities $\rho^{\chi}_{\text{br}}$ (second panels) and $\rho^{\chi}_{\text{rad}}$ (fourth panels) on the model parameters $\nu$ and $\lambda$. The daughter coupling as been set to $g=10^{-3}$ for all simulations. Dots represent the numerically-computed timescales in the two-fields scenario. Dashed lines refer to the fitting formulas obtained from Eq.~\eqref{eq:fit_rho_br} and \eqref{eq:fit_rho_rad} while solid lines are derived from the lattice simulations of the single-spectator-field scenario. Lighter lines correspond to smaller quartic self-couplings (down to $\lambda=10^{-5}$) while darker lines to larger couplings, up to $\lambda=10^{-1}$.}
    \label{fig:z_timescales_total_daughter}
\end{figure}

\begin{figure}[tb]
    \centering
    \includegraphics[width=0.9\textwidth]{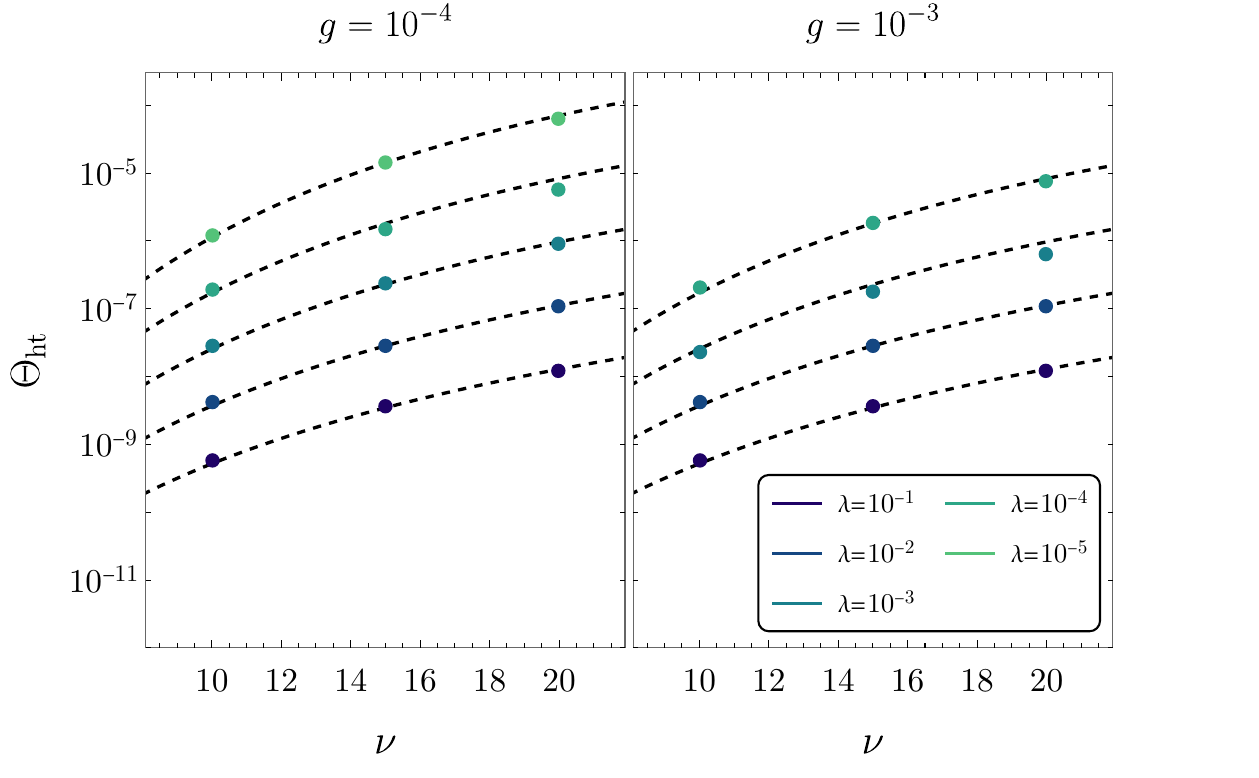}
    \caption{Parametric dependence of the heating efficiency $\Theta_{\text{ht}}$ on the two model parameters $\lambda$ and $\nu$. Dashed lines correspond to the fitting formulas obtained for the single-scalar-field case. The colour coding matches the one defined for the previous figures.}    \label{fig:efficiency_temp_daughter}
\end{figure}

The parametric dependence of the timescales $z_{\text{br}}$, $z_{\text{rad}}$ and of the energy densities $\rho^{\chi}_{\text{br}}$, $\rho^{\chi}_{\text{rad}}$ following from our simulations is presented in Fig.~\ref{fig:z_timescales_total_daughter}. We observe that, for the values of $g$ considered in our simulations, the end of the tachyonic instability is caused primarily by the self-interactions of the spectator scalar field, thus matching closely the results obtained in the single field case. After the peak production of the spectator field modes, the associated kinetic and gradient energies evolve as radiation, with both fields contributing to the overall virialisation of the system. The timescale $z_{\text{rad}}$ is computed numerically by extending the condition in~\eqref{eq:vir_criterium_lattice} to a 2-component interacting fluid, i.e. by including the additional kinetic, gradient and interaction terms. Again, we notice a good matching between $z_{\text{rad}}$ for the single- and two-fields scenarios. The interval between $z_{\text{rad}}$ and $z_{\text{eq}}$ has a variable duration, depending on many factors, including the intensity of the production processes, the appearance of parametric resonances and the magnitude of the coupling constant $g$. For the cases $(g=10^{-3}, \lambda=10^{-3})$ and $(g=10^{-4}, \lambda=10^{-4})$, the two fields are reaching equipartition of the total energy already during the virialisation phase. This phenomenon induces stronger oscillations on the spectator field which makes the estimation of $z_{\text{rad}}$ somewhat less reliable, see the middle panel of Fig.~\ref{fig:z_timescales_total_daughter}. On the other hand, a short equipartition timescale allows us to compute it numerically for the special cases $(g=10^{-3}, \lambda=10^{-3})$ and $(g=10^{-4}, \lambda=10^{-4})$, finding it to be in the range $250<z_{\text{eq}}<400$. After rescattering between fields has redistributed the total energy density, the system evolves towards the end of the heating phase. We notice once more that the results in Fig.~\ref{fig:z_timescales_total_daughter} are in excellent agreement with the previous results for the a single spectator field. Lastly, Fig.~\ref{fig:efficiency_temp_daughter} displays the heating efficiency for the two-field setup with $g=10^{-3}$ and $g=10^{-4}$ compared with the parametric formula obtained from Eq.~\eqref{eq:fittingTheta}.

It is interesting to notice that for sufficiently large couplings $g \sim \mathcal{O}(0.1)$ and small self-interactions $\lambda$, the backreaction is expected to be produced by the daughter field itself. In this scenario, the rapid growth of $\varphi$ fluctuations could dynamically generate an effective mass for the spectator field at large times, even if the quartic self-coupling is completely negligible ($\lambda \to 0$). Unfortunately, the energy transfer to UV modes in this case is fast and affects a large band of momenta that cannot be properly covered by a lattice with $N=512$, the largest one we are currently able to simulate. Therefore, we leave the testing of this enticing scenario to a future work.

\section{Conclusions and outlook}\label{sec:conclusion}
 
In the present article, we have carried out an extensive analysis of the Hubble-induced heating mechanism appearing in quintessential inflation scenarios involving spectator fields non-minimally coupled to gravity. By performing a large number of 3+1-dimensional classical lattice simulations, we have probed a wide range of the parameter space of the theory, numerically determining the main timescales in the process, namely the initial tachyonic instability, the onset of backreaction effects and the virialisation phase leading to a radiation-like equation-of-state for the spectator field. These findings allowed us to derive a set of parametric equations that can be conveniently applied to evaluate the efficiency of the heating process, the temperature at the onset of radiation domination and the minimum number of e-folds of inflation required to solve the flatness and horizon problems in specific quintessential scenarios, without relying on further lattice simulations. In this sense, our results constitute a major step forward in addressing the question of heating in non-oscillatory models of inflation 
\cite{Felder:1999pv,Rubio:2017gty,Dimopoulos:2017zvq,Dimopoulos:2017tud,BuenoSanchez:2007jxm}, within a minimal, natural and non-perturbative setup. In particular,  since the gravitational coupling considered in this work is required by the renormalisation of the energy-momentum tensor in curved space-time \cite{Birrell:1982ix,Callan:1970ze}, our findings provide a lower bound on the heating efficiency and the associated radiation temperature, allowing generically for successful big bang nucleosynthesis even in the absence of direct couplings between the inflaton and the matter sector.

For the sake of completeness, we also explored a prototypical yet significant extension of the single-field scenario. By introducing an explicit coupling with a secondary scalar sector beyond the standard model, we have assessed the potential depletion of the spectator field fluctuations prior to the onset of non-linearities. The two-field scenario constitutes an excellent testing ground for the robustness of the single-scalar-field fitting formulas. Indeed, our analysis shows that, for moderate values of the coupling between fields, the results of the single-field model constitute an accurate description of the two-fields model as well. Note also that these results can be directly applied to scenarios involving Abelian gauge bosons as daughter fields, as first shown in Ref.~\cite{Figueroa:2015rqa} in the context of parametric resonance and explicitly verified here.  

Our approach could be easily adapted to scenarios far beyond the toy-model cases considered in this work. The most direct application would be the Standard Model itself, with a Higgs field stable up to the Planck scale \cite{Bezrukov:2014ina} playing the role of the spectator field.~\footnote{For works assuming the \textit{instability} of the electroweak vacuum at high energies, see e.g. Refs.~\cite{Figueroa:2016dsc, Figueroa:2017slm, Opferkuch:2019zbd}.} In this context, the analysis of secondary interactions becomes extremely important, since the tachyonic productions of Higgs perturbations is expected to compete with the secondary amplification of gauge bosons and their potential decay into fermions, along the lines of the \textit{Combined Preheating} scenario developed in Refs.~\cite{Garcia-Bellido:2008ycs,Repond:2016sol,Fan:2021otj}. Interestingly enough, the inclusion of these dissipative decay channels could strongly modify the single-field picture presented in this paper, opening the door to stabilising the defects generated during the transition and modifying with it the associated gravitational waves spectrum \cite{Bettoni:2018pbl,Kamada:2015iga}. 

Further applications of our setting could include models with a larger field content, involving, for instance, additional dark matter or beyond-the-standard sectors, as well as higher-dimensional operators. In this context, it would be interesting to explore the early non-thermal production of dark matter relics \cite{Fairbairn:2018bsw, Laulumaa:2020pqi, Babichev:2020xeg} and the generation of the matter-antimatter asymmetry in the Universe \cite{Bettoni:2018utf}. Finally, directly coupled gauge fields, involving for instance Chern-Simons interactions, could also be considered as interesting secondary fields with a rich phenomenology, e.g. in the context of primordial magnetogenesis. 
 

\acknowledgments 
We thank Dario Bettoni and Asier Lopez-Eiguren for useful discussions. We also thank Toby Opferkuch for the useful discussion on the latest version of the manuscript. G.L. (ORCID 0000-0002-4739-4946) is supported by a fellowship from ”la Caixa” Foundation (ID 100010434) with fellowship code LCF/BQ/DI21/11860024. G.~L. thanks also the Funda\c c\~ao para a Ci\^encia e a Tecnologia (FCT), Portugal, for the financial support to the Center for Astrophysics and Gravitation-CENTRA, Instituto Superior T\'ecnico,  Universidade de Lisboa, through the Project No.~UIDB/00099/2020. J.R. (ORCID ID 0000-0001-7545-1533) is supported by a Ram\'on y Cajal contract of the Spanish Ministry of Science and Innovation with Ref.~RYC2020-028870-I. The authors acknowledge the Department of Theoretical Physics of the Universidad Complutense de Madrid for allowing access to the cluster facilities utilised for the research carried out in this paper.
 
\bibliographystyle{ieeetr}
\bibliography{bib}

\end{document}